\documentclass[showpacs,aps,pre,twocolumn,amsmath,amssymb,superscriptaddress]{revtex4-1}
\usepackage{graphicx}
\usepackage{dcolumn}
\usepackage{bm}
\usepackage{gensymb}
\usepackage{mathcomp}
\usepackage{textcomp}
\usepackage{setspace}

\usepackage{color}

\begin{document}
\title{Differential Colorimetry Measurements of Fluctuation Growth
in Nanofilms Exposed to Large Surface Thermal Gradients}

\author{Kevin R. Fiedler}
\altaffiliation{Current address: Dept. of Mathematics and Statistics, Washington State University, 2710 Crimson Way, Richland, WA 99354}
\affiliation{California Institute of Technology, T. J. Watson Sr. Laboratories of Applied Physics, \\
MC 128-95, Pasadena, CA 91125, USA}
\author{Euan McLeod}
\affiliation{College of Optical Sciences, University of Arizona,\\
1630 E. University Blvd. P.O. Box 210094, Tucson, AZ 85721, USA}
\author{Sandra M. Troian}
\email[Corresponding author: ]{stroian@caltech.edu}
\homepage[URL: ]{http://www.troian.caltech.edu}
\affiliation{California Institute of Technology, T. J. Watson Sr. Laboratories of Applied Physics, \\
MC 128-95, Pasadena, CA 91125, USA}

\date{\today}

\begin{abstract}
Slender liquid nanofilms exposed to large surface thermal gradients are known to undergo thickness fluctuations which rapidly self-organize into arrays of nanoprotrusions with a separation distance of tens of microns. We previously reported good agreement between measurements of the characteristic spacing and the wavelength of the most unstable mode predicted by a linear stability analysis based on a long wavelength thermocapillary model. Here we focus on differential colorimetry measurements to quantify early time out-of-plane growth of protrusions for peak heights spanning 20 to 200 nm. Analysis of peak heights based on shape reconstruction reveals robust exponential growth. Good quantitative agreement of the growth rates with the thermocapillary model is obtained using a single fit constant to account for material parameters of nanofilms that could not be measured directly. These findings lend further support to the conjecture that the array protrusions uncovered almost two decades ago stem from a linear instability whose growth rate is controlled by thermocapillary forces counterbalanced by capillary forces.
\end{abstract}

\maketitle

\section{Background}
Experiments designed to elicit the physical mechanisms generating linear instabilities in macroscale fluid systems normally rely on early time measurements of an emergent length or time scale signalling the growth of the most unstable wavelength. This is common to measurements in many large scale systems which manifest stationary periodic, oscillatory uniform or oscillatory periodic phenomena. At the macroscale, such measurements can often be obtained by direct visualization. With growing interest in small scale fluidic phenomena, similar measurements pose more challenges - not only do films easily rupture or are otherwise compromised by defects but measurements must often rely on indirect techniques from which key parameter value are inferred. At microscale or nanoscale dimensions, matching system size and materials properties to the appropriate measurement technique often severely restricts the options available. For films whose thickness is of the order of tens of microns or more, laser or white light interferometry, sometimes coupled with shadowgraphy, is often the tool of choice \cite{VanHook:prl1995,VanHook:jfm1997,Garnier:prl2003}.

For more than a decade now, there has been interest in identifying the source of various  runaway instabilities in even thinner films, which undergo spontaneous transition from an initially flat and uniform layer to microarrays containing nanoprotusions. In these systems, fluid elongations tend to grow without bound unless prematurely terminated by direct contact with an opposing substrate or by fluid depletion effects. Early time measurements of the fastest growing wavelength in polymeric nanofilms subject either to large surface electric field gradients \cite{Chou:jvstb1999,Zhuang:Princeton2002,Schaffer:nat2000,Leach:macro2005} or large surface thermal gradients \cite{Dietzel:prl2009,Dietzel:jap2010,McLeod:prl2011,Fiedler:jap2016} have yielded a characteristic in-plane separation distance of the order of tens of microns. Current understanding of these systems is that capillary forces, which suppress development of regions of high interfacial curvature, are counterbalanced and then eventually dominated either by electrohydrodynamic or thermocapillary forces which rapidly undergo self-reinforcement, leading to fluid elongations with long range order.

Experimentalists investigating the electrohydrodynamic instability have had some success in measuring out-of-plane growth for film thicknesses of about 5 $\mu$m or less. Leach, Lin and Russell \cite{Leach:macro2005,russell:book2009} used laser scanning confocal microscopy with single wavelength illumination (458 nm) to obtain real time measurements of the film deformation process. The field of view spanned roughly 750 by 750 $\mu \textrm{m}^2$ and encompassed about 10 to 20 liquid peaks. Frame by frame analysis of the fringe spacing associated with the first peak to touch the counter-electrode allowed reconstruction of the evolving shapes, which seemed well described by a Gaussian function. The measured growth in peak heights was found to be consistent with exponential growth as predicted by a linear stability analysis \cite{Schaffer:nat2000,Schaffer:epl2001}, although the number of data points was rather small and the scatter was significant. These measurements revealed how the growth rate was strongly influenced by the applied voltage difference, the initial film thickness and the liquid viscosity. In a separate study, Leach \textit{et al.} also used scanning electron microscopy to capture still images of film deformation which accompanied  hierarchical formation in trilayer systems consisting of two polymeric films (total thickness of about 500 nm) overlaid by an air layer \cite{Leach:chaos2005}.

In this work, we focus on early time measurements of the out-of-plane growth of nanoprotrusions triggered by thermal gradients in much thinner films. This initial period of growth poses significant experimental challenges since emergent fluctuations in film thickness measure less than a few tens of nanometers. We therefore resort to a technique based on differential colorimetry applied to surface reflectivity images of ultrathin transparent films. While the method in principle is well suited to measurement of liquid peak growth in nanoscale films, the material films themselves pose problems because they are prone to defects from dust and other contaminants. These defects, which can be embedded or exposed, tend to trigger localized nucleation and growth which interfere with measurements of the native instability. Gathering statistically relevant data from a sufficiently large region of a film not compromised by such defects proves a challenging task.

In what follows, we report the first measurements of the early time growth rate of fluctuations spanning 20 to 200 nm in height in nanofilms exposed to large surface thermal gradients. These measurements rely on frame by frame tracking of the ten most rapidly growing liquid peaks which are monitored and analyzed using differential colorimetry \cite{Hartl:opteng1997}. The experimental design allowed a fairly large field of view (about 1.36 \textrm{mm} x 1.08 $\textrm{mm}$) which contained about 100 liquid peaks per frame. Reconstruction of peak heights at early times revealed robust exponential growth. The measured growth rates were found to compare well quantitatively to predictions of a long wavelength thermocapillary model \cite{Dietzel:prl2009,Dietzel:jap2010,McLeod:prl2011,Fiedler:jap2016} using a single adjustable parameter to account for material parameters of nanofilms that could not be measured directly.

\section{Modal Growth Rate Prediction From Long Wavelength Thermocapillary Model}
\label{TCModel}
\begin{figure}
\centering
\includegraphics[scale=0.8]{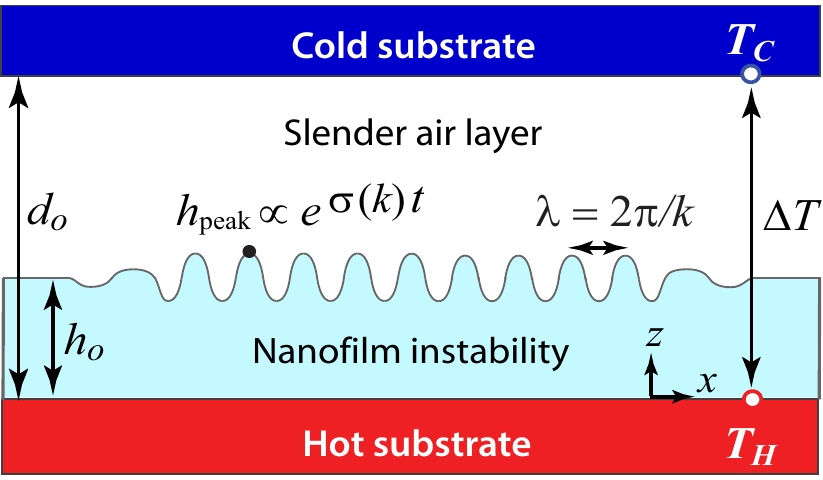}
\caption{(Color online) Sketch of an unstable liquid nanofilm of initial thickness $h_o$  sandwiched between two substrates separated by a small distance $d_o$ and maintained at a uniform temperature difference $\Delta T = T_\text{H} - T_\text{C}$, where \text{H} and \text{C} designate hot and cold, respectively. Rapidly growing unstable modes with wave number $k = 2 \pi / \lambda$ give rise to liquid peak amplitudes $h_{peak} \propto  e^{\sigma (k) t}$, where $\sigma (k)$ is given by Eq. (\ref{eq:dispersion}).}
\label{fig:theorysketch}
\end{figure}

Dietzel and Troian \cite{Dietzel:prl2009,Dietzel:jap2010} previously analyzed the linear instability of a slender fluid bilayer consisting of a Newtonian liquid nanofilm overlaid by an air or gas layer and exposed to a large transverse thermal gradient. A simplified sketch depicting an unstable liquid film with periodicity $\lambda$ and peak height $h_{\textrm{peak}} \propto e^{\sigma(k)t}$ is shown in Fig. \ref{fig:theorysketch}. The instability arises whenever an ultrathin liquid film with initial uniform thickness $h_o$ is placed between two solid substrates separated by a small distance $d_o$ (typically less than a couple microns) and maintained at a uniform temperature difference $\Delta T = T_\text{H} - T_\text{C}$, where \text{H} and \text{C} designate hot and cold, respectively. The slender air layer above the liquid nanofilm is treated as a passive inviscid and poorly conducting film with thermal conductivity $k_{\textrm{air}}$. The liquid layer thermal conductivity is given by $k_\textrm{liquid}$. In typical experiments using polymeric films, $\kappa = k_\textrm{air}/k_\textrm{liquid} \approx 1/4$. The long wavelength model (based on the slender gap geometry) predicates an aspect ratio $(d_o/\lambda)^2 << 1$ and large gradients in surface temperature along the air/liquid interface due to transverse thermal conduction. Thermal convection and radiation are  estimated to be orders of magnitude smaller than transverse conduction \cite{Dietzel:jap2010,Fiedler:jap2016}. Stabilizing gravitational forces are also assumed to be negligible. Additional details of the model including the full derivation leading to the final expressions for the wavelength and growth rate of the fastest growing mode can be found in Ref. [\onlinecite{Dietzel:jap2010}].
Dirichlet thermal boundary conditions imposed on the hot and cold substrate enforce heat transfer by thermal conduction across the slender bilayer. The validity of the model is restricted to $0 < h(\vec{x},t) \leq d_o$ where $h(\vec{x},t)$ denotes the local film thickness at position $\vec{x}$ at time $t$. At early times, infinitesimal spontaneous modal fluctuations in film thickness can be described by
\begin{equation}
h(\vec{x},t) = h_o + \delta h \, e^{\sigma(k) t} \, e^{i \vec{k} \cdot \vec{x}} \, ,
\label{eq:fluctuation}
\end{equation}
where $\delta h << h_o$, $k=|\vec{k}(x,y)| = 2 \pi\lambda$ and $\sigma(k)$ is real. Given the small gap size $d_o$, these fluctuations give rise to significant variation in surface temperature, which in turn generate large thermocapillary forces along the air/liquid interface defined by $z=h(\vec{x},t)$. These forces, mitigated only by capillary forces, lead to development of nanoprotrusions which rapidly advance toward the cooler substrate. Under the  assumptions of this model, the dispersion curve obeys a Type II instability \cite{Cross:Cambridge2009} of the form:
\begin{equation}
\sigma(k) = \frac{h_o k^2}{\eta} \left[ \frac{\kappa \,  (d\gamma/dT)\, D \, \Delta T}{2(D + \kappa - 1)^2}\, - \frac{\gamma \, h_o^2}{3}\, k^2\right],
\label{eq:dispersion}
\end{equation}
where $D = d_o/h_o$ is the normalized separation ratio, $\kappa$
is the thermal conductivity ratio and  $d\gamma/dT$ is the thermocapillary coefficient. For very small amplitude fluctuations characteristic of early time growth, the material parameters $\eta$, $k_{\textrm{liquid}}$, $\gamma$ and $d\gamma/dT$ are treated as constants evaluated at an appropriate reference temperature. It is common to choose as the reference temperature either the temperature corresponding to the original flat interface at $z=h_o$, or more typically, the temperature of the supporting substrate  $T_\text{H}$. These two temperatures are normally very close in magnitude since the temperature drop across the liquid film is typically very small because the majority of the temperature drop occurs within the air film due to its lower thermal conductivity.

The extremum of Eq. (\ref{eq:dispersion}) yields an expression for the wavelength of the fastest growth mode $k_o =|\vec{k}_o (x,y)|$ given by
\begin{equation}
k_o  = \frac{2 \pi}{\lambda_o} = \frac{1}{h_o} \sqrt{ \frac{3 \kappa (d\gamma/dT) \Delta T}{4 \gamma} } \, \frac{\sqrt{D}}{(D + \kappa - 1)}
\label{eq:kmax}
\end{equation}
whose growth rate $\sigma_o$ is given by
\begin{equation}
\sigma_o = \frac{3 \, \kappa^2 (d\gamma/dT)^2 \, (\Delta T)^2}{16 \, \eta \, \gamma \, h_o} \, \frac{D^2}{\left(D + \kappa - 1\right)^4} \, .
\label{eq:sigmamax}
\end{equation}
For later purposes, we here also introduce the relation between the non-dimensional maximal growth rate $\Sigma_o$ and the normalized wavelength $\Lambda_o = \lambda_o/h_o$ for the fastest growing mode:
\begin{equation}
\Sigma_o = \frac{\eta \, h_o }{\gamma} \sigma_o  = \frac{1}{3} \left (\frac{2 \pi}{\Lambda_o} \right )^4 \, .
\label{eq:SIGMA}
\end{equation}
This relation holds for any Type-II instability where the dispersion curve $\sigma(k)$ consists of a positive quadratic term dominant at small wave numbers $k$ counteracted by a negative quartic term dominant at large wave numbers. For the long wavelength instability of interest, the quadratic term represents the destabilizing thermocapillary forces and the quartic term represents stabilizing capillary forces.

\section{Brief Review of Experimental System}
\begin{figure}
\centering
\includegraphics[scale = 0.8]{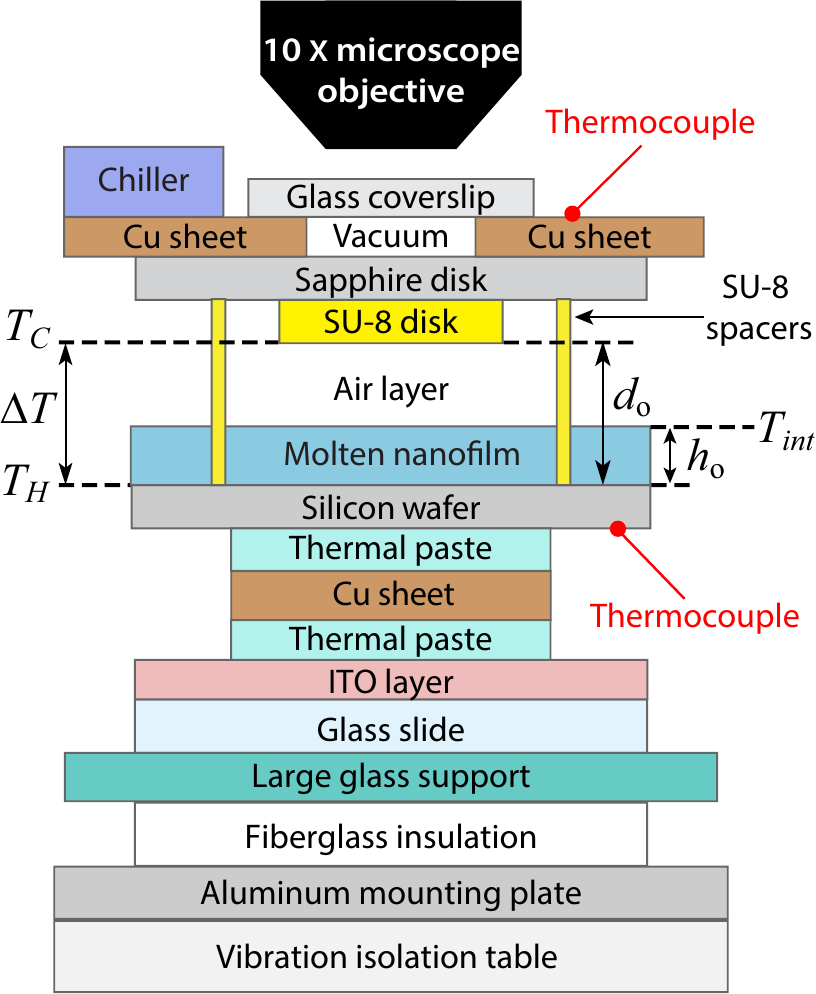}
\caption{(Color online) Diagram of the geometry and material layers comprising the experimental system (not drawn to scale). Exact geometric dimensions, material constants and other relevant information for each layer can be found in Ref. \onlinecite{Fiedler:jap2016}. Relevant ranges of experimental values can be found in  Table~\ref{table:params}.}
\label{fig:expsystem}
\end{figure}

Details of the experimental system, film preparation techniques and finite element simulations for estimating $\Delta T$ are described elsewhere \cite{McLeod:prl2011,Fiedler:jap2016}. Here we briefly review key elements of the layered structure depicted in Fig.\ref{fig:expsystem} comprising the experimental setup. The molten nanofilms used in these studies were formed of polystyrene (1.3 kg/mol PS standard, M$_w$/M$_n$ = 1.10, Scientific Polymer Products, Inc.) dissolved in toluene and filtered to remove any contaminants, undissolved or agglomerated material [0.2 $\mu$m PTFE filter (Cole Palmer), 200 nm nominal pore size]. The solutions were then spun coat onto the polished side of a silicon wafer to the desired thickness (CEE-100, Brewer Science). The wafers were then vacuum baked at 80 \degree C for two hours to evaporate  residual solvent (Precision Model 19, Thermo Fisher Scientific). The final film thickness, $h_o$, measured in the solid phase by ellipsometry (Gaertner Model L166C), ranged from about 95 to 390 nm. To begin an experimental run, the chiller was turned on and the PS coated wafer was positioned where shown in Fig.\ref{fig:expsystem}. Power was applied to the indium tin oxide (ITO) coated glass slide to initiate heating and melting of the PS film. Once the temperature of the hot wafer reached steady state, which normally required from 1.5 to 5 minutes, data collection began. All the images analyzed in this study (e.g. Fig. \ref{fig:peakGrowth}) were obtained well after the temperature had stabilized so that the temperatures $T_H$ and $T_C$ remained constant during the course of measurement.

Digital images showing the growth of fluctuations were obtained with a color CCD camera (DVC 1312C) mounted onto the porthole of a Zeiss Axiovert 200 MAT microscope at 10X magnification. A halogen light (100 W max., Osram, HLX 64625) provided the free space illumination. Images were captured every one to two minutes by a frame grabber card (PIXCI-D) and image capture software (XCAP, Epix Inc.). All the interstitial layers situated between the silicon wafer and glass coverslip, including the SU-8 photoresist disk (when in place), were optically transparent to the halogen light.

In order to obtain estimates of the instability growth rate, values of the temperature difference $\Delta T = T_H - T_C$ are required inputs to Eq. (\ref{eq:sigmamax}). Direct measurement of this quantity has always been problematic in all experiments to date since the separation distance $d_o$ (665 to 2200 nm in this study) is too small to accommodate even the smallest of thermocouples. Other techniques like thermal imagery lack the resolution needed to provide reliable estimates of $\Delta T(\vec{x},t)$. A review of how the quantity $\Delta T(\vec{x},t)$ was obtained in previous experiments described in Refs. [\onlinecite{Chou:jvstb1999,Zhuang:Princeton2002,Schaffer:am2003,Schaffer:epl2002,
Schaffer:am2003,Schaffer:Konstanz2001}] indicates that the values reported were the set point values for the hot and cold stages and not the actual temperature drops across the confined air/liquid bilayer. If indeed the case, then the values that have been reported are likely overestimates of $\Delta T$. In our experiments, we confirmed that measurements based on the set points obtained from the two thermocouple positions shown in Fig. \ref{fig:expsystem}) yielded large overestimates of $\Delta T$. Such  overestimates are due to the fact that the interstitial layers situated between the thermal source (ITO coated glass slide) and sink (chiller), however thin, cause non-negligible thermal series resistances, which ultimately degrade the temperature drop across the air/liquid bilayer. Furthermore, the many edges of these interstitial layers act like thermal fins which induce further heat loss.  As described in more detail in Ref. \onlinecite{Fiedler:jap2016}, our estimates of $T_\text{H}$ and $T_\text{C}$ within the gap region extrated from numerical simulations were always calibrated against actual thermocouple readings at locations exterior to the gap. The simulated values reported in Table \ref{table:master} were substituted into Eq. (\ref{eq:sigmamax}) to obtain improved theoretical estimates of instability growth rates which were then compared to the experimental measurements.

The chromatic analysis described in the next section was used to investigate 20 of the 26 experiments originally reported in Ref. \onlinecite{Fiedler:jap2016}. The same numbering convention originally used to tag experimental runs was used here as well for consistency. For various reasons, the six discarded runs did not undergo sufficient growth to be analyzed reliably by the protocol used. Relevant variables for the 20 acceptable experimental runs are provided in Tables \ref{table:params}, \ref{table:layerthickness} and \ref{table:master}.

The remainder of this work is organized as follows. In Section \ref{sec:colorimetry}, we describe the procedures and algorithms based on differential colorimetry used to quantify the growth of liquid peak heights of the fastest growing unstable modes. In Section \ref{sec:comparison}, we compare results of these measurements to predictions of the growth rates obtained from a linear instability analysis of a long wavelength thermocapillary model and find good agreement using a single fit parameter. In Section \ref{sec:discussion}, we discuss limitations of this current study and suggest experimental improvements anticipated to yield even closer agreement with theory.

\begin{table}
\begin{small}
\caption{Dimensions and material constants for the experiments reported in this study. Nanofilms used were polystyrene melts (1.3 kg/mol PS standard, M$_w$/M$_n$ = 1.10, Scientific Polymer Products, Inc.). Values of the surface tension were estimated from the literature by extrapolation in molecular weight and temperature \cite{Pu:polydatabook1999}. Viscosities were inferred by extrapolation in temperature from the zero shear measurements reported by Urakawa \textit{et al.} \cite{Urakawa:macro2004} for 1.8 kg/mol PS, the molecular weight closest to our samples. Additional information regarding the values listed below can be found in Ref. [\onlinecite{McLeod:prl2011}] and Ref. [\onlinecite{Fiedler:jap2016}].}
\label{table:params}
\vspace{0.05in}
\def\arraystretch{1.1}
\begin{ruledtabular}
\begin{tabular}{l c r}
\textbf{Parameter}  & \textbf{Symbol} & \textbf{Value}  \\
\hline
Initial film thickness (nm)              & $h_o$     & 95-390 \\
Substrate separation distance (nm)       & $d_o$     & 615-2200 \\
Normalized separation distance $d_o/h_o$ & $D$       & 1.97-14.1 \\
Liquid/solid interface temp. (\degree C) & $T_H$  & 89.8-101 \\
Air/solid interface temp. (\degree C)    & $T_C$  & 67.8-88.7 \\
Temperature difference $T_H - T_C$ (\degree C)& $\Delta T$ & 9.83-25.8\\
Liquid surface tension (100 \degree C) & $\gamma$   & 35.5 \\
~~~~~(mN/m)  (est. from Ref. \onlinecite{Pu:polydatabook1999}) &&\\
Thermocapillary coeff. ($\mu$N/m--\degree C)& $|d\gamma/dT|$ &  78 \\
~~~~~(est. from Ref. \onlinecite{Pu:polydatabook1999}) &&\\
Thermal conductivity ratio & $\kappa$ & 0.2422 \\
~~~~~$k_\textrm{air}/k_\textrm{liquid}$&&\\
(est. from Refs. \onlinecite{Welty:Wiley1984,Pu:polydatabook1999})&&\\
PS viscosity at $T_H$ (Pa-s) & $\eta (T_H)$ & 31-94 \\
~~~~~(est. from Ref. \onlinecite{Urakawa:macro2004}) &&\\
PS viscosity at $T_C$ (Pa-s) & $\eta (T_C)$   & 105-3930 \\
~~~~~(est. from Ref. \onlinecite{Urakawa:macro2004}) &&\\
Fastest growing wavelength ($\mu$m) & $\lambda_o$& 29.1-73.2 \\
~~~~~(measured from expt.)&&\\
Instability growth rate ($10^{-4}$/s) & $\sigma_o$ & 0.345-158 \\
~~~~~(measured from expt.)&&\\
\end{tabular}
\end{ruledtabular}
\end{small}
\end{table}

\section{Differential Colorimetry Measurements for Quantifying Peak Fluctuation Amplitudes
\label{sec:colorimetry}}

\subsection{Total reflectance of multilayer stack}
Differential colorimetry offers a very useful tool for quantifying variations in chromatic interference fringes that arise from spatiotemporal variations in film thickness. The reconstruction method for extracting film shapes eliminates problems often incurred by use of other techniques since human observers perceive colors differently. The technique relies on direct comparison of the RGB (red-green-blue) values of each pixel in a digital image to the predicted RGB colors based on reflectance information from a multilayer stack that includes the deforming liquid film as well the spectral response of the illumination source and the camera used to record the image \cite{Hartl:opteng1997}. For the system of interest, a multilayer stack was used to compute the (ideal) fraction of incident light reflected from the air/polymer interface as a function of wavelength. Given the relatively large distance between the glass coverslip and the microscope objective, it was assumed that the illumination was normally incident on the planar stack. The ideal or theoretical fractions of RGB for a given film thickness were obtained by convolving the reflection spectrum with the illumination and camera spectral functions. The RGB values from pixels of interest obtained from experimental images were then compared to the theoretical curves and the fit optimized via a cost function.

Given that the liquid peak heights protruding above the initial flat film value $h_o$ were typically just a few nanometers, it was difficult to identify the fastest growing modes unambiguously at early stages of the formation process. This problem was resolved by carrying out the peak reconstruction process in inverse chronological order since the  peaks of interest were clearly identifiable at late times after they had undergone more growth. A similar approach was successfully implemented in Ref.~[\onlinecite{Fiedler:jap2016}] for the purpose of extracting the wavelength of the fastest growing mode. For each experimental run characterized by $h_o$, $d_o$ and $\Delta T$, it was relatively simple to identify the maximum height of the ten fastest growing peaks from image sequences taken at fixed intervals $\Delta t$ in time, where $\Delta t$ ranged from ten seconds to two minutes depending on the control settings governing each run.

The ideal total reflectance was obtained by assuming white light normally incident on a multilayer stack consisting of seven layers arranged in the following sequence - glass, vacuum, sapphire, SU-8, air, polystyrene and silicon - as indicated in Fig. \ref{fig:expsystem}. (For experiments 66 and 69 in Table \ref{table:master} in which the SU-8 disk was not used, the stack consisted of only six multilayers.) The values of the refractive index as a function of illumination wavelength $\lambda_\text{illum}$ were obtained from the Cauchy equation \cite{Jenkins:1976} given by
\begin{equation}\label{eq:Cauchy}
n(\lambda_\text{illum}) = B + \frac{C}{\lambda_\text{illum}^2} + \frac{D}{\lambda_\text{illum}^4} ~.
\end{equation}
The Cauchy coefficients are listed in Table~\ref{table:Cauchy}. Silicon has both real and imaginary Cauchy coefficients since the solid absorbs light in the visible part of the spectrum. The refractive index for sapphire was chosen to correspond to the ordinary axis since the orientation of the extraordinary axis in experiment was unknown. The layer thicknesses in the multilayer stack are listed in Table \ref{table:master} and Table \ref{table:layerthickness}.  For each material layer, the Cauchy coefficients were substituted into Eq.~(\ref{eq:Cauchy}) and the refractive index value $n(\lambda_\text{illum})$ computed in increments of 1 nm over the  range 0.4 $\mu\text{m} \leq \lambda_\text{illum} \leq 0.8~\mu \text{m}$. Any unknown interstitial values were obtained by linear interpolation from the two closest values.

\begin{table}
\begin{small}
\centering
\caption{Cauchy coefficients for evaluating refractive indices as a function of illumination wavelength $\lambda_\text{illum}$ ($\mu$m) within the visible spectrum for the material layers shown in Fig. \ref{fig:expsystem}.}
\label{table:Cauchy}
\def\arraystretch{1.5}
\begin{ruledtabular}
\begin{tabular}{l c c c }
Material & $B$ & $C$  &  $D$\\
&     & $10^{-2}\, \mu \text{m}^2$  & $10^{-4} \,\mu \text{m}^4$\\
\hline
Corning 1737 glass [\onlinecite{Corning}]   	& 1.505    & 0.455      	& -0.218  	 \\
Vacuum or air (est)   	& 1.000    & 0.000      	& 0.000  	\\
Sapphire (ord axis) [\onlinecite{Tydex}]	& 1.750    & 0.654      	& -1.31  	 \\
SU-8 [\onlinecite{MicroChem}]				& 1.566    & 0.796     		& 1.40  	\\
Polystyrene [\onlinecite{Nikolov:apop2000}]	& 1.563    & 0.929       	& 1.20  	\\
Silicon (real) [\onlinecite{Green:prog1995}]  & 3.819    & -17.2       	 & 727  	\\
Silicon (imag) [\onlinecite{Green:prog1995}] & 0.106    & -8.14       	& 167  	\\
\end{tabular}
\end{ruledtabular}
\end{small}
\end{table}

Ideal total reflectance values $R$ for the multilayer stack were computed using the matrix formulation method \cite{Yeh:2005}. This compact formalism yields a simple 2$\times$2 matrix description for plane wave reflection, propagation and transmission through a stack consisting of layers of isotropic homogenous material. The matrix product yields the total transfer matrix whose elements are used to compute the total reflectance values at normal incidence.

In the analysis below, the planar interfaces are designated by $1 \leq N \leq 6$ for integer $N$, and the individual material layers denoted by $0 \leq j \leq N+1$, where the integer $j=1$ represents the glass coverslip layer and $j = 6$ represents the silicon wafer. The semi-infinite air layer above the glass coverslip corresponds to $j = 0$. For simplicity, the silicon wafer was assumed to be bounded below by a semi-infinite air layer defined by $j = 7$. The Fresnel reflection and transmission coefficients for an interface separating layers $j$ and $j+1$ are then given by \cite{Born:1999}
\begin{equation} \label{eq:rij}
r_{j,j+1} = \frac{n_j-n_{j+1}}{n_j + n_{j+1}}
\end{equation}
\begin{equation} \label{eq:tij}
t_{j,j+1} = \frac{2n_j}{n_j + n_{j+1}} \, .
\end{equation}

The 2$\times$2 matrix for transmission across an interface separating layer $j$ from layer $j+1$ is given by \cite{Yeh:2005}:
\begin{equation} \label{eq:layer_trans}
\mathbf{T}_{j,j+1} = \frac{1}{t_{j,j+1}}
\begin{bmatrix}
 1  & r_{j,j+1} \\
 r_{j,j+1}  & 1
\end{bmatrix} \, .
\end{equation}
Similarly, the matrix describing propagation through layer $j$ is given by \cite{Yeh:2005}
\begin{equation} \label{eq:layer_prop}
\mathbf{T}_{j} =
\begin{bmatrix}
 e^{-i \delta_j}  & 0 \\
 0  & e^{-i \delta_j}
\end{bmatrix} \, ,
\end{equation}
where the phase $\delta_j$ obtained when passing through layer $j$ of
thickness $z_j$ is given by
\begin{equation} \label{eq:deltai}
\delta_j(\lambda_\text{illum}) = \frac{2 \pi n_j}{\lambda_\text{illum}} \, z_j \, .
\end{equation}
The total transfer matrix $\mathbf{M}$ was then computed according to
\begin{equation} \label{eq:transfer_matrix}
\mathbf{M} =
\begin{bmatrix}
M_{11}  & M_{12} \\
M_{21}  & M_{22}
\end{bmatrix} = \mathbf{T}_{0,1}\mathbf{T}_{1}\mathbf{T}_{1,2}\mathbf{T}_{2} \dots \mathbf{T}_{N}\mathbf{T}_{N,N+1}~.
\end{equation}
The value of the total reflectance \cite{Yeh:2005} for pixel location $(\vec{x},t)$ in a layered stack illuminated by wavelength $\lambda_\text{illum}$ for a local film thickness $h(\vec{x},t)$ was then obtained according to
\begin{equation} \label{eq:reflectance}
R \left [ \lambda_\text{illum}, h(\vec{x},t) \right ] = \frac{|M_{21}|^2}{|M_{11}|^2} \, .
\end{equation}
These values were then inverted to reconstruct the shape and
maximum amplitudes of the fastest growing liquid peaks as a function of time.

\begin{table}
\begin{small}
\centering
\caption{Layer thicknesses for glass coverslip, sapphire window and silicon wafer depicted in Fig. \ref{fig:expsystem}.}
\label{table:layerthickness}
\def\arraystretch{1.5}
\begin{ruledtabular}
\begin{tabular}{l c }
Layer material & Thickness (\text{mm}) \\
\hline
Corning 1737 glass coverslip & 0.150\\
Cylindrical sapphire window & 0.400\\
Silicon wafer & 0.675
\end{tabular}
\end{ruledtabular}
\end{small}
\end{table}

\subsection{Color maps for extracting fluctuation growth}
\begin{figure}
\includegraphics[scale=0.8]{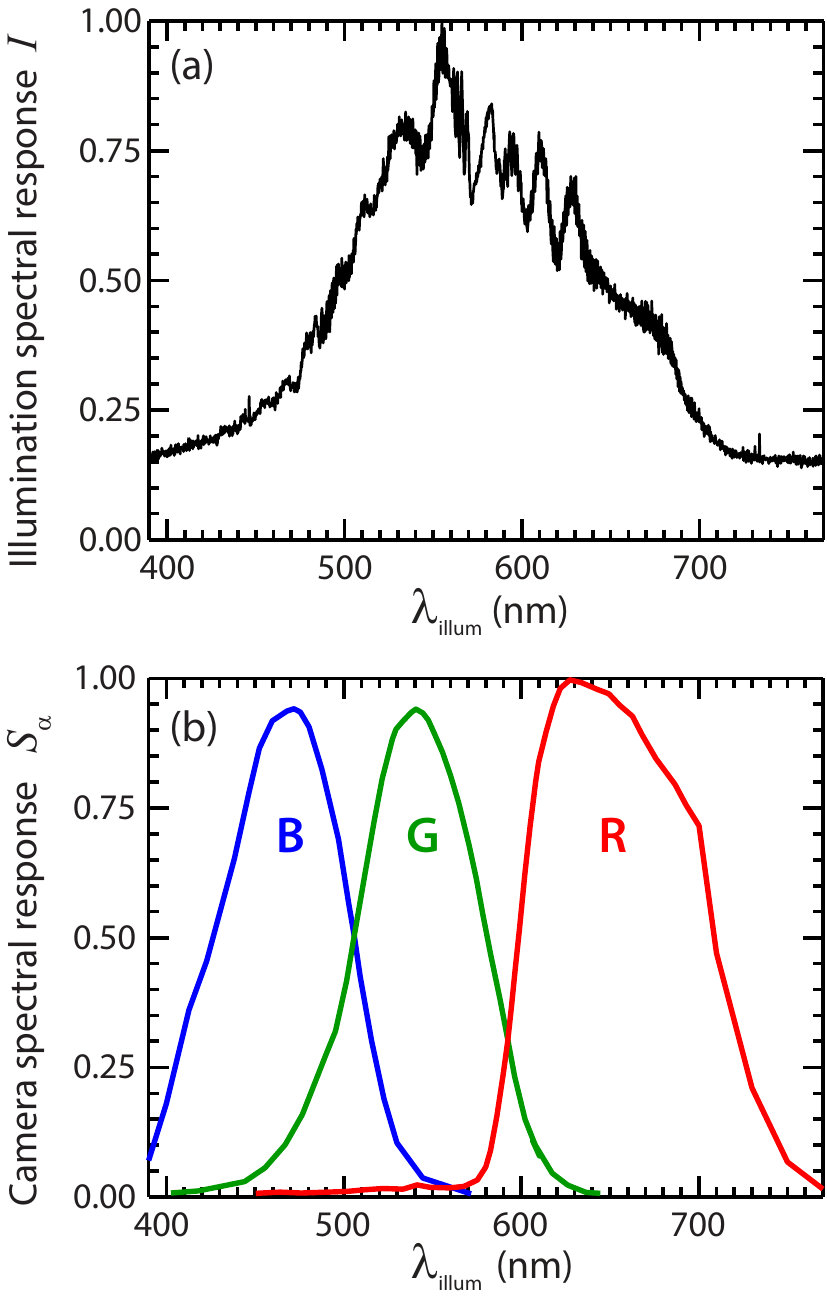}
\caption{(Color online) (a) Measured spectral response $I(\lambda_\text{illum})$ of halogen light illumination used in this study (Osram HLX 64625, 12 V, 100 W). (b) Digital camera (DVC 1312C) spectral response functions $S_\alpha (\lambda_\text{illum})$ for the individual red (R), green (G) and blue (B) channels as provided by the manufacturer.}
\label{fig:spectralplots}
\end{figure}

\begin{figure}
\centering
\includegraphics[scale=0.9]{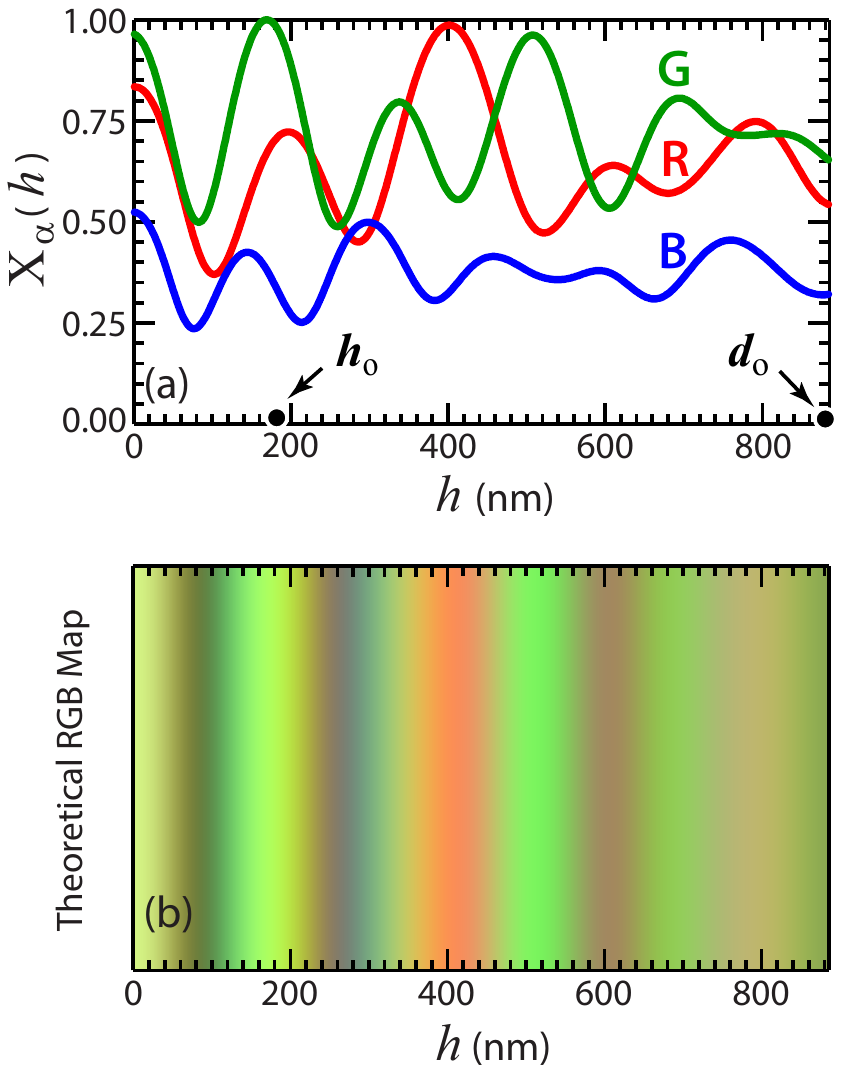}
\caption{(Color online) (a) Tristimulus curves computed according to Eq.~(\ref{eq:Xalpha}) as a function of liquid film thickness $h$ for experimental run 56. (Additional input values can be found in Table \ref{table:layerthickness} and Table \ref{table:master}.) (b) Theoretical RGB color map computed as described in the text as a function of liquid film thickness $h$.}
\label{fig:interf}
\end{figure}

As the linear instability began sprouting protrusions, the recorded image displayed  variations in local color due to changes in surface reflectance associated with the changing surface topology of the liquid nanofilm. We sought to quantify the growth rate of the fastest growing modes by examining the RGB content of pixels in the vicinity of emergent liquid peaks. To derive the color cap for an \textit{ideal} multilayer stack, the tristimulus values for each color channel $\alpha$ were computed from the convolution integral $\text{X}_\alpha(h)$ where $\alpha = 1 \rightarrow$ red (R), $\alpha = 2 \rightarrow$ green (G) and $\alpha = 3 \rightarrow$ blue (B) according to
\begin{equation}
\hspace{-0.8pt}
\text{X}_\alpha(h) = \int I(\lambda_\text{illum}) R(\lambda_\text{illum}, h)
S_\alpha (\lambda_\text{illum}) \; d\lambda_\text{illum} \, ,
\label{eq:Xalpha}
\end{equation}
where $I(\lambda_\text{illum})$ denotes the spectral response of the illumination source and $S_\alpha (\lambda_\text{illum})$ denotes the spectral response of the digital camera. Fig.~\ref{fig:spectralplots}(a) depicts the normalized function $I(\lambda_\text{illum})$ for the halogen source used in this study (Osram HLX 64625, 12 V, 100 W), which was measured by placing a spectrometer (Ocean Optics USB4000-VIS-NIR) at the focal plane of the nanofilm. The three curves $S_\alpha (\lambda_\text{illum})$ shown in Fig.~\ref{fig:spectralplots}(b) were obtained from the camera manufacturer (DVC 1312C). Shown in Fig. \ref{fig:interf}(a) are the tristimulus curves computed from Eq.~(\ref{eq:Xalpha}) as a function of local film thickness $h$ for the parameter values corresponding to experimental run 56 (see Table \ref{table:layerthickness} and Table \ref{table:master}). Each curve was normalized by its maximum value which for the blue channel was outside the range shown. The theoretical RGB map shown in Fig.~\ref{fig:interf}(b) was computed by summing the R, G and B component values for each value of $h$, in increments of 1 nm. Similar maps were generated for each experimental run.

There color maps were also adjusted to account for automatic changes changes in image brightness and white balance that occurred from one image to another in a given sequence. These changes were enforced through internal camera settings not accessible to users.  Image brightness settings, which vary with the exposure time, affected the values of the minima and maxima computed from Eq.(\ref{eq:Xalpha}). The optimal white balance settings set by the camera, internally adjusted by modifying the relative weights of each color channel, also changed from run to run since the actual illumination conditions varied somewhat from  experiment to experiment. To account for these effects, we therefore normalized the maximum and minimum values computed from Eq.(\ref{eq:Xalpha}) for each color channel $\alpha$ independently by linearly stretching the curves until the two extrema matched the R, G and B values extracted from experiment. This transformation is given by
\begin{eqnarray}
\min_{h \in [h_o, h_{\text{final}}]} \widehat{\text{X}}_\alpha [h(\vec{x},t)] &=& \min_{t \in [0,t_\text{final}]} \text{X}^\text{exp}_\alpha (\vec{x}_\text{final}, t) \nonumber\\
\max_{h \in [h_o, h_{\text{final}}]} \widehat{\text{X}}_\alpha [h(\vec{x},t)] &=& \max_{t \in [0,t_\text{final}]} \text{X}^\text{exp}_\alpha (\vec{x}_\text{final}, t) \, ,\nonumber\\
\label{eq:adjust}
\end{eqnarray}
where $\vec{x}_\text{final}$ denotes the Cartesian coordinate of a pixel analyzed at the final time $t_\text{final}$ of an experimental run, $h_o$ denotes the local layer thickness at the initial time and $h_\text{final}$ denotes the layer thickness corresponding to a local maximum peak height achieved at time $t_\text{final}$. For those runs in which the selected protrusions made contact with the cold substrate, the values $h_\text{final}$ were set equal to $d_o$. However, for those cases in which the gap spacing $d_o$ was very large and the liquid did not make contact with the cold substrate (i.e. typically in the absence of the SU-8 disk), the values $h_\text{final}$ were  estimated by matching the experimentally observed color of the selected peaks to the theoretically computed color.

\subsection{Extraction of liquid peak growth rates}
Estimates of the maximum amplitudes $h_\text{peak}(t)$ for the ten fastest growing liquid protrusions were obtained as follows. A cost or penalty function $G(\vec{x}, t, h)$ was defined according to
\begin{equation}
G(\vec{x},t;h) = \sum_{\alpha=1}^3 \left [\text{X}^\text{exp}_\alpha (\vec{x}, t) - \widehat{\text{X}}_\alpha [h(\vec{x},t)] \right]^2
\label{eq:cost}
\end{equation}
to minimize the difference between the theoretical and experimental RGB values. Due to the oscillatory nature of the curves shown in Fig. \ref{fig:interf}(a) as well as uncertainties in film thickness arising from noisy data or small lateral variations in thickness of the flats comprising the multilayer stack, it was not always possible to find the global minimum of $G(\vec{x}, t, h)$ within the range $h_o \leq h \leq h_\text{final}$. This was often the case at early times when liquid peak heights had only grown by a few nanometers. For this reason, the image analysis was conducted in reverse chronological order, using the locations and colors of well defined peaks at late times. Starting from a selected peak at the final time, the next numerical search for the peak height explored a neighborhood containing film thicknesses not more than 70 nm below the peak height value just obtained and so on. Additionally, the algorithm allowed the Cartesian coordinates of the peak being analyzed, i.e. $[x_\text{peak}(t),~y_\text{peak}(t)]$, to shift between frames by one pixel to the right or left and one pixel up or down. In this way, film elevations for the five pixels within the neighborhood of the peak analyzed in the previous frame were easily computed. This relaxed search scheme accommodated very small lateral shifts in peak locations which occurred from image to image, either due to spurious effects or small lateral flow caused by uneven substrates or nearby spacers. The final value accepted, $h_\text{meas}(x, y, t)$, was chosen to be the value within the neighborhood of a peak which minimized the cost function $G(\vec{x}, t, h)$ subject to the constraint
$h \in [h_\text{peak} (t + \Delta t) - 70 \;\text{nm}, \;\; h_\text{peak} (t + \Delta t)]$. The values $h_\text{peak}(t)$ and $[x_\text{peak}(t), \; y_\text{peak}(t)]$ so computed identified the location and peak heights at time $t$ for the ten fastest growing peaks.

\begin{figure}
\centering
\includegraphics[scale=0.8]{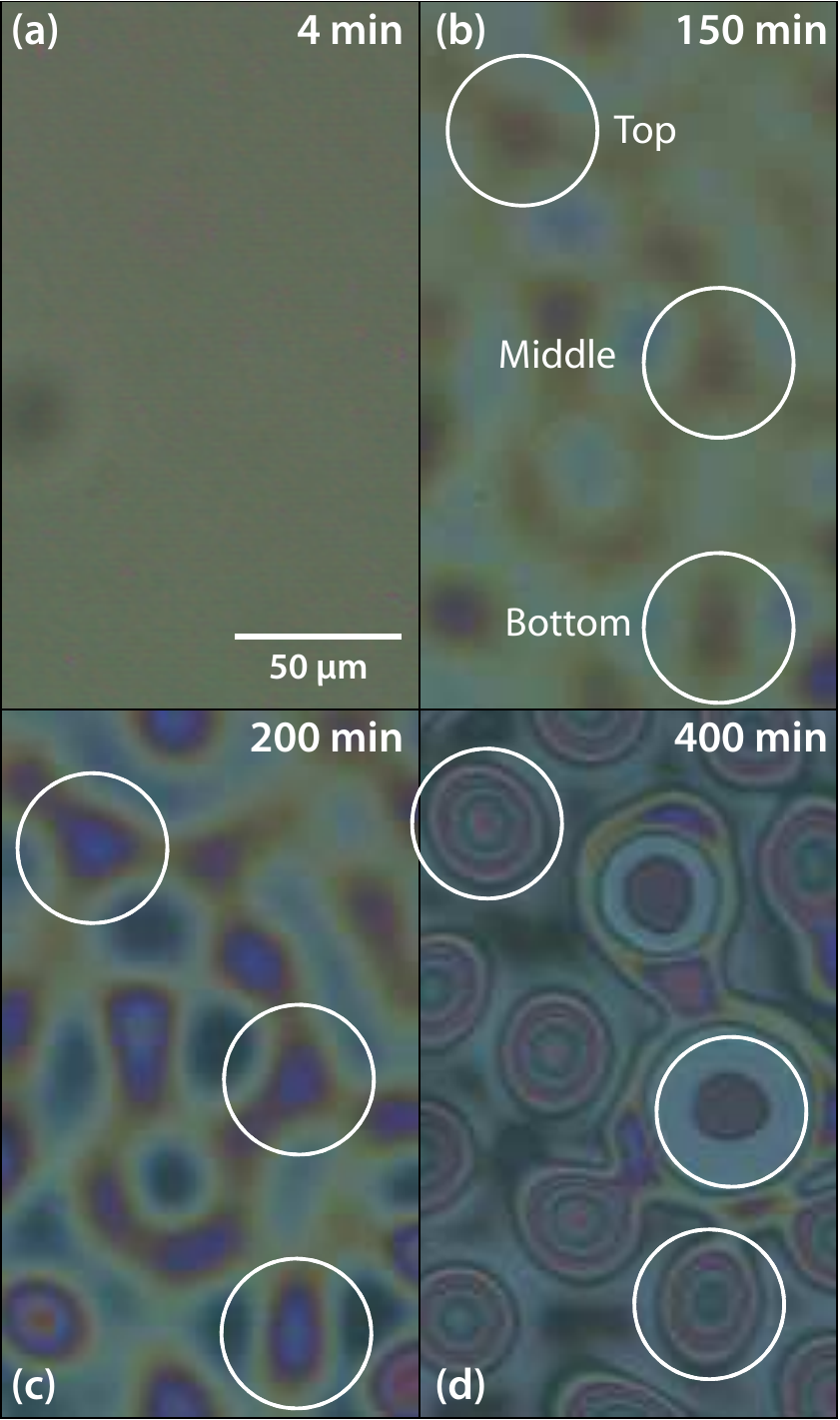}
\caption{(Color online) Four bright field micrographs from experimental run 56 (see Table \ref{table:master}) showing the same region of the molten PS film at $t$ (min) = 4, 150, 200 and 400 as it underwent instability. The white circles designate three of the ten fastest growing peaks, labeled Top, Middle and Bottom for later reference.}
\label{fig:circledpeaks}
\end{figure}

\begin{figure}
\begin{center}
\includegraphics[scale=0.9]{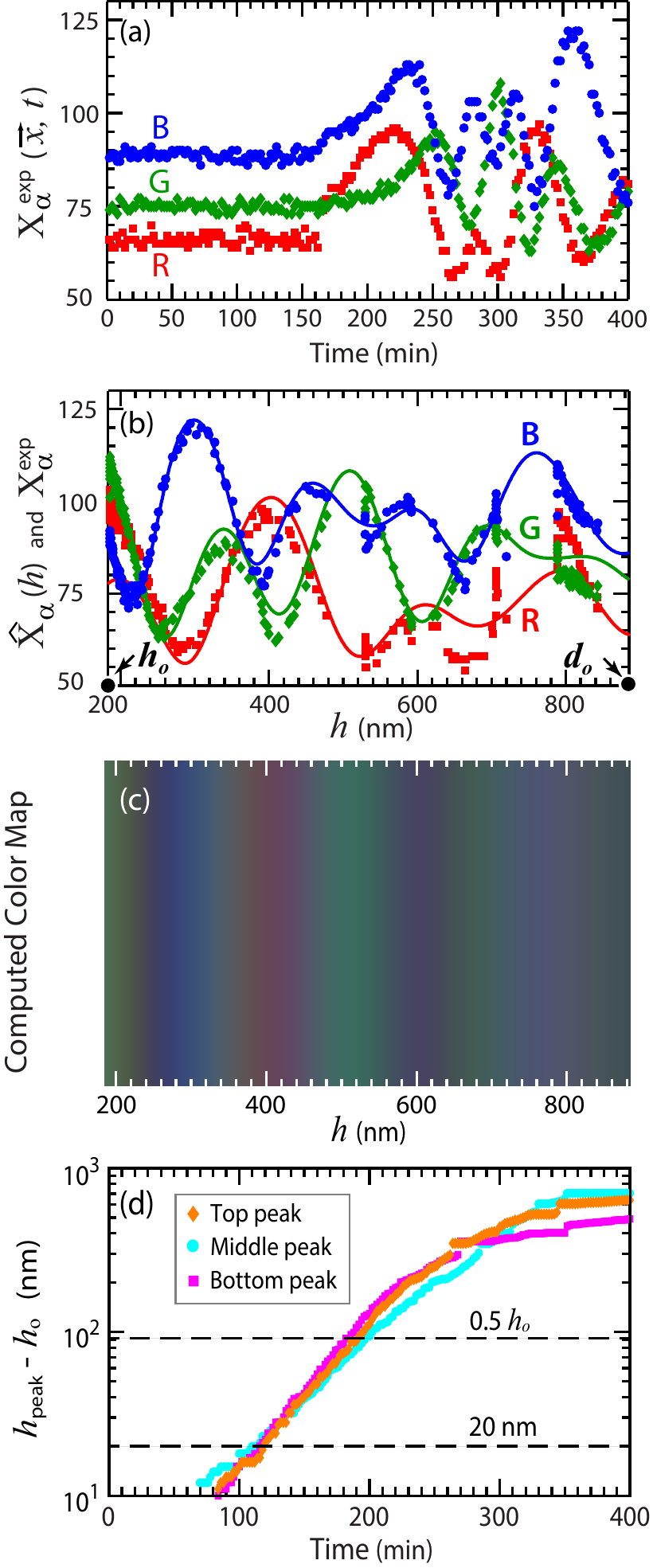}
\caption{(Color online) (a) Measured RGB values as a function of time,
$X_\alpha ^\textrm{exp}(\vec{x},t)$, for the peak labeled Top in Fig. \ref{fig:circledpeaks}. (b) Match between the tristimulus values extracted from the peak heights in experiment (icons), $X_\alpha ^\textrm{exp}(\vec{x},t)$, with those computed from Eq. (\ref{eq:Xalpha}) and rescaled according to Eq. (\ref{eq:adjust}) (solid lines) as obtained from minimization of the cost function in Eq. (\ref{eq:cost}). (c) Color map corresponding to the RGB values given by the solid lines in (b). (d) Reconstruction of the peak height values as a function of time for the three peaks circled in Fig.~\ref{fig:circledpeaks}. Linear fits to the curves (not shown) between the limits of 20 nm and $h_o/2$ were used to extract values of the maximum growth rate $\sigma^{\text{exp}}$.}
\label{fig:peakGrowth}
\end{center}
\end{figure}

This procedure is demonstrated next for experiment 56 (see Table \ref{table:master}). Four snapshots of the emergent instability are shown in Fig.~\ref{fig:circledpeaks}. These images depict the same region of the PS film with initial thickness $h_o = 183$ nm and $\Delta T = 22.1\degree \textrm{C}$ at times $t$ (min) = 4, 150, 200 and 400. Three of the ten fastest growing peaks are circumscribed by white circles. The RGB values of each peak as a function of time were extracted from the peak centroids. Fig. \ref{fig:peakGrowth}(a) shows the measured RGB values as a function of time for the peak labeled Top in Fig. \ref{fig:circledpeaks}. The maximum and minimum values of each color curve in Fig. \ref{fig:peakGrowth}(a) were then used to rescale the corresponding theoretical tristimulus curves computed from Eq. (\ref{eq:Xalpha}). Fig.~\ref{fig:peakGrowth}(b) depicts the resulting match obtained from minimization of the cost function in Eq. (\ref{eq:cost}) between the RGB values extracted experimentally from pixel locations representing peak heights (icons) to those RGB values computed theoretically (solid lines) from the stretched tristimulus curves. Plotting the RGB values for each estimated layer thickness $h$ yielded the color map shown in  Fig.~\ref{fig:peakGrowth}(c). As evident, the rescaling (i.e. stretching) procedure described above led to a color map which much more closely resembles the colors observed in experiment - see Fig. \ref{fig:circledpeaks}. In general, this matching procedure worked well for layer thickness values below about $1.5 h_o$. Occasionally, it led to small jumps in the reconstructed heights in the vicinity of points where $\widehat{\text{X}}_\alpha [h(\vec{x},t)]$ reached an extremum. An example of this is evident in Fig.~\ref{fig:peakGrowth}(c), near the values $h$ = 500 nm and $h$ = 750 nm, where both the green and red channels reach an extremum at similar values of $h$. The fitted curves helped bridge the behavior between those discontinuous values in film thickness.

Shown in Fig.~\ref{fig:peakGrowth}(d) are the resultant semi-log plots for the reconstructed peak heights $h_\text{peak} - h_o$ for the three protrusions highlighted in Fig.~\ref{fig:circledpeaks}. Values of the maximum growth rate $\sigma_o^{\text{exp}}$ were extracted from the linear portions of these curves occurring between 20 nm and $h_o/2$ - these bounding values were used to analyze all runs. This linear response is evidence of exponential growth, as predicted by Eq. (\ref{eq:fluctuation}), although this behavior typically persisted only for one decade in time or less. In all cases examined, initial exponential growth gave way to slower than exponential growth at intermediate and late times. This subsequent slower growth was likely caused by local film depletion and  rapid increase in viscosity as the liquid peaks advanced toward the colder substrate. The viscosity of polymer melts are known to follow Arrhenius-like dependence on temperature, $e^{E/RT}$, where $E$ is an activation energy and $R$ the universal gas constant.

The field of view of images analyzed in this way was fairly large (1.36 mm $\times$  1.08 mm) and typically encompassed about 100 protrusions. Only the ten fastest growing peaks were selected for analysis and their corresponding growth rates reported in Table \ref{table:AllGrowthRates} for the 20 experiments listed in Table \ref{table:master}. While the wave numbers associated with these ten peaks did not necessarily all correspond to the unique wave number $k_o$ characterizing the fastest growing mode, we assumed nonetheless that their wavenumbers were closely clustered about that value. This assumption is a reasonable one given that the dispersion curve for the instability is governed by Type-II behavior \cite{Cross:Cambridge2009} and as described by Eq. \ref{eq:dispersion} exhibits a well defined peak about the value $k_o$. For each experimental run, we therefore averaged the growth rates of the ten fastest growing peaks and used that averaged value $\sigma_o^{\textrm{ave}}$ in reporting the maximum growth rate for each run analyzed. These average values are listed in the last column of Table \ref{table:master}.

\section{Comparison of Measured to Predicted Growth Rates From Thermocapillary Model
\label{sec:comparison}}
As a first test, we compared the measured growth rates $\sigma_o$ to the theoretical values estimated from Eq. (\ref{eq:SIGMA}). Since linear stability analysis derives from the early time response of a system to infinitesimal perturbations, it was assumed that the appropriate reference value to use for the fluid viscosity $\eta$ in Eq. (\ref{eq:SIGMA}) was $\eta_{T_H}$, the value corresponding to the temperature $T_H$ of the initial flat film. At the final time of growth, the local viscosity of liquid peaks is closer to $\eta_{T_C}$, where $T_C$ is the temperature of the cold substrate. These two limiting values therefore establish the bounds on $\eta$ within the confined gap. Listed in Table \ref{table:master} are the corresponding values of the viscosities $\eta_{T_H}$ and $\eta_{T_C}$ as estimated by linear interpolation in temperature from a series of measurements made by Urakawa \textit{et al.} \cite{Urakawa:macro2004} for the range in temperature 62 \degree C to 216 \degree C. The polystyrene used in our study was obtained from the same supplier but was of slightly lower molecular weight ($M_w = 1.3$ kg/mol instead of $M_w = 1.9$ kg/mol), too small a difference to affect the viscosity in any significant way. in fact, we found good agreement between Urakawa \textit{et al.}'s reported value of the glass transition temperature, $T_g = 60$ \degree C, and the temperatures at which we noted film softening. Given that polymer viscosity increases exponentially with decreasing temperature while surface tension increases only linearly so, we used reference values $\gamma = \gamma(T_H)$ in our analysis as well - those  values are listed in Table \ref{table:master}.

\begin{figure}
\centering
\includegraphics[scale=0.9]{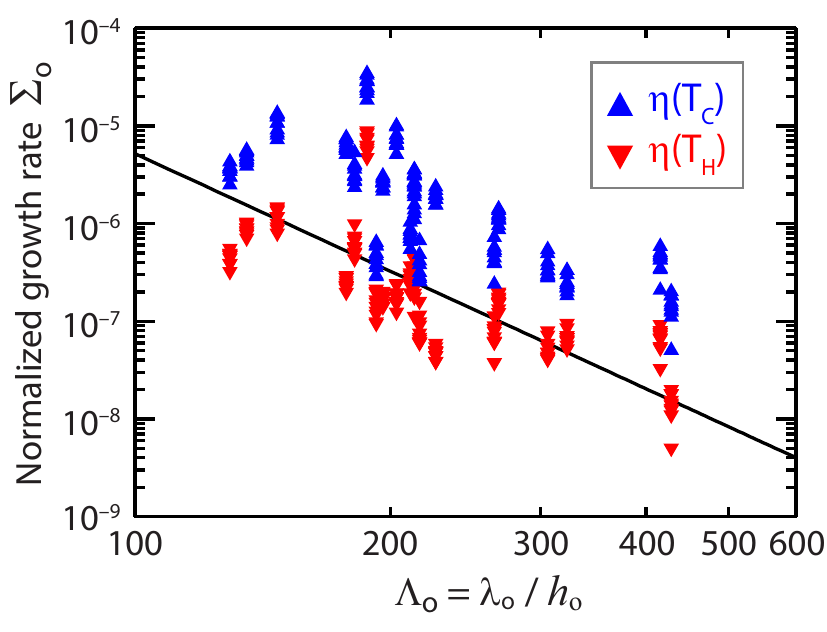}
\caption{(Color online) Correlation between the normalized growth rate and normalized wavelength of the fastest growing mode. Triangles (blue): Normalized growth rates $\eta(T_H)h_o \sigma_o/\gamma(T_H)$. Inverted triangles (red): Normalized growth rates $\eta(T_C)h_o \sigma_o/\gamma(T_H)$. Solid line (black): Eq.~(\ref{eq:SIGMA}).}
\label{fig:SIGMALAMBDA}
\end{figure}

Shown in Fig. \ref{fig:SIGMALAMBDA} is a comparison of the normalized growth rates $\Sigma_o = \eta h_o \sigma_o/\gamma$ versus the normalized instability wavelength $\Lambda_o = \lambda_o/h_o$ for the ten fastest growing peaks observed for each experimental run listed in Table \ref{table:master} using the upper and lower bounds on viscosity. Superimposed on the data is the prediction
$\Sigma_o = 1/3 (2 \pi /\Lambda_o)^4$ given by Eq. (\ref{eq:SIGMA}). The correlation observed is consistent with the prediction of the linear stability theory based on the thermocapillary model. The normalized growth rates computed with $\eta(T_H)$ agree more closely with the theoretical prediction, which is encouraging since significant effort was made in this study to extract growth rates at early times when the film temperature was close to $T_H$.

\begin{figure}
\centering
\includegraphics[scale=0.9]{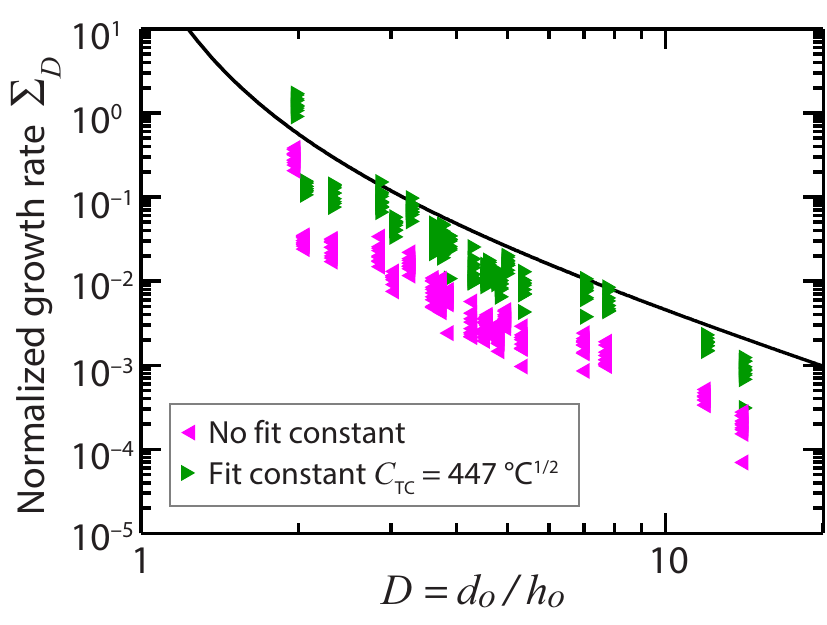}
\caption{Normalized instability growth rates $\Sigma_D$ versus normalized separation distance $D=d_o/h_o$ given by Eqs. (\ref{eq:normsigmamax}) and (\ref{eq:SIGMADADJCONST}). The solid curve represents the function $D^2/3(D + \kappa - 1)^4$.}
\label{fig:BThermocap}
\end{figure}

We also examined the correlation between the growth rate of the fastest growing mode and the normalized separation distance $d_o/h_o$, as predicted by Eq.(\ref{eq:sigmamax}). In recent work, we reported \cite{Fiedler:jap2016} that a single adjustable parameter, $C_{\text{TC}}$, allowed good quantitative agreement to be obtained between experimental measurements of the fastest growing wavelength $\lambda_o$ and predictions based on Eq. (\ref{eq:kmax}). This fit constant was determined to be $C_{\text{TC}} = 447~( \degree C)^{1/2}$, about 45\% larger than the estimate derived from the theoretical expression for that constant, namely $C_{\text{TC}} = 2 \pi (4 \gamma)^{1/2}/ [3 \kappa (d\gamma/dT)]^{1/2}$. We reported that this discrepancy was likely due to uncertainties in the values of the material constants, whose values we extracted from the literature since direct measurement was not possible. Eq.(\ref{eq:sigmamax}) was recast according to
\begin{equation}
\Sigma_D =\frac{16 \, \eta(T_H)\, \gamma (T_H) \, h_o}{9 \, \kappa^2 (d\gamma/dT)^2 \, (\Delta T)^2} \, \sigma_o = \frac{D^2}{3 \left(D + \kappa - 1\right)^4} \, ,
\label{eq:normsigmamax}
\end{equation}
to contain no adjustable fit parameter or equivalently
\begin{equation}
\Sigma_D=\frac{\eta(T_H) \, h_o \, (C_{\text{TC}})^4}{(2 \pi)^4 \gamma(T_H) (\Delta T)^2} \, \sigma_o = \frac{D^2}{3 \left(D + \kappa - 1\right)^4} \, ,
\label{eq:SIGMADADJCONST}
\end{equation}
which incorporates a single adjustable parameter $C_{\text{TC}} = 447~( \degree C)^{1/2}$.
Plotted in Fig. \ref{fig:BThermocap} are the measured values of $\Sigma_D$, with and without the fit constant, along with the function $D^2/3(D + \kappa - 1)^4$. A single fit constant yields good quantitative agreement between theory and experiment.

\begin{table*}[!]
\begin{small}
\begin{singlespace}
\caption{Parameter values for all experiments reported in this study. Dimensions and symbols refer to the system geometry depicted in Fig. \ref{fig:expsystem} - additional layer thicknesses are listed in Table \ref{table:layerthickness}. Columns below denote the following: Expt. is the number assigned each run using the same numbering convention in Ref.~[\onlinecite{Fiedler:jap2016}]; $d_o$ is separation distance between hot and cold substrates; $h_o =$ is the initial thickness of the PS film; $D = d_o/h_o$ is the normalized separation distance; SU-8 is the thickness of the SU-8 disk; $\lambda_o$ is the wavelength of fastest growing unstable mode, as measured in Ref.~[\onlinecite{Fiedler:jap2016}]; $T_\text{H}$ and $T_\text{C}$ are the temperatures of the (hot) PS /silicon and (cold) air/SU 8 interfaces  (or air/sapphire interface when the SU 8 disk was absent) as computed from finite element simulations \cite{Fiedler:jap2016}; $\Delta T = T_H - T_C$; $\eta(T_\text{H})$ and $\eta(T_\text{C})$ are the viscosity of PS estimated from Fig. 7 of Ref. \onlinecite{Urakawa:macro2004}; $\sigma^{\textrm{min}}_o$), $\sigma^{\textrm{max}}_o$, and $\sigma^{\textrm{ave}}_o$ denote the minimum, maximum and average values of the measured growth rates for the ten fastest growing peaks listed in Table V.} \label{table:master}
\end{singlespace}
\begin{ruledtabular}
\begin{tabular}{c | c | c | c | c | c | c | c | c | c | c | c | c | c | c }
\multicolumn{1}{c}{Expt.} & \multicolumn{1}{c}{$d_o$} & \multicolumn{1}{c}{$h_o$}
& \multicolumn{1}{c}{$D$} & \multicolumn{1}{c}{SU-8} &\multicolumn{1}{c}{$\lambda_o$}
& \multicolumn{1}{c}{$T_\text{H}$} & \multicolumn{1}{c}{$T_\text{Int}$}
& \multicolumn{1}{c}{$T_\text{C}$} & \multicolumn{1}{c}{$\Delta T$} & \multicolumn{1}{c}{$\eta(T_\text{H})$} &
\multicolumn{1}{c}{$\eta(T_\text{C})$} & \multicolumn{1}{c}{$\sigma^{\textrm{min}}_o$} & \multicolumn{1}{c}{$\sigma^{\textrm{max}}_o$} & \multicolumn{1}{c}{$\sigma^{\textrm{avg}}_o$} \\
\multicolumn{1}{c}{} & \multicolumn{1}{c}{(nm)} & \multicolumn{1}{c}{(nm)}
& \multicolumn{1}{c}{} & \multicolumn{1}{c}{($\mu$m)}
&\multicolumn{1}{c}{($\mu$m)} & \multicolumn{1}{c}{(\degree C)}
&\multicolumn{1}{c}{(\degree C)} & \multicolumn{1}{c}{(\degree C)} & \multicolumn{1}{c}{(\degree C)} & \multicolumn{1}{c}{(Pa-s)}
&\multicolumn{1}{c}{(Pa-s)} & \multicolumn{1}{c}{($10^{-4}$/s)} & \multicolumn{1}{c}{($10^{-4}$/s)} & \multicolumn{1}{c}{($10^{-4}$/s)} \\
\hline
46 & 900 & 189 & 4.76 & 1.38 & 37.0 & 93.6 & 92.4 & 73.8 & 19.8 & 60.6 & 925 & 4.11 & 5.89 & 5.03\\

47 & 810 & 189 & 4.29 & 1.38 & 40.9 & 99.0 & 98.1 & 86.3 & 12.7 & 36.1 & 152 & 2.94 & 7.79 & 4.18\\

48 & 840 & 183 & 4.59 & 1.38 & 32.5 & 89.9 & 88.6 & 69.3 & 20.6 & 93.9 & 2440 & 3.84 & 5.68 & 4.78\\

50 & 900 & 181 & 4.97 & 1.38 & 48.5 & 96.8 & 95.8 & 80.8 & 16.0 & 45.2 & 323 & 4.99 & 8.13 & 6.57\\

51 & 700 & 181 & 3.87 & 1.48 & 38.6 & 91.8 & 90.5 & 75.4 & 16.4 & 76.5 & 755 & 2.70 & 8.74 & 7.01\\

52 & 605 & 185 & 3.27 & 1.48 & 39.1 & 98.5 & 97.6 & 88.7 & 9.83 & 37.2 & 105 & 9.45 & 17.9 & 13.1\\

53 & 830 & 183 & 4.54 & 1.38 & 59.1 & 100 & 99.7 & 88.1 & 12.4 & 33.0 & 116 & 2.87 &
5.23 & 3.79\\

54 & 670 & 185 & 3.62 & 1.48 & 35.6 & 99.0 & 98.1 & 88.4 & 10.6 & 36.2 & 109 & 4.73 & 10.7 & 7.53\\

56 & 885 & 183 & 4.84 & 1.38 & 37.2 & 89.8 & 88.6 & 67.8 & 22.1 & 94.4 & 3930 & 2.39 & 4.68 & 3.50\\

58 & 760 & 142 & 5.35 & 1.48 & 37.6 & 96.1 & 95.3 & 81.1 & 15.0 & 48.3 & 310 & 1.83 & 5.51 & 3.60\\

60 & 735 & 95.0 & 7.74 & 1.48 & 29.1 & 95.9 & 95.4 & 80.4 & 15.5 & 49.3 & 337 & 2.91 & 5.65 & 3.79\\

61 & 710 & 101 & 7.03 & 1.48 & 42.0 & 95.8 & 95.2 & 80.9 & 14.9 & 50.0 & 319 & 2.16 & 6.10 & 4.41\\

62 & 785 & 258 & 3.04 & 1.48 & 33.4 & 93.9 & 92.2 & 78.2 & 15.7 & 59.3 & 461 & 7.03 & 12.1 & 9.83\\

63 & 754 & 201 & 3.75 & 1.48 & 42.9 & 94.1 & 92.9 & 78.7 & 15.5 & 58.0 & 413 & 5.09 & 12.8 & 8.65\\

64 & 615 & 215 & 2.86 & 1.48 & 39.0 & 93.1 & 91.6 & 80.1 & 13.0 & 64.6 & 351 & 10.4 & 23.8 & 15.1\\

66 & 2200 & 156 & 14.1 & 0.00 & 66.8 & 101 & 101 & 81.0 & 20.3 & 30.9 & 314 & 0.345 & 1.38 & 0.974\\

69 & 2200 & 184 & 12.0 & 0.00 & 41.6 & 97.2 & 96.6 & 71.4 & 25.8 & 43.2 & 1740 & 1.62 & 2.49 & 1.99\\

71 & 850 & 366 & 2.32 & 1.48 & 53.8 & 93.7 & 91.3 & 77.4 & 16.4 & 59.9 & 549 & 12.1 & 22.5 & 17.7\\

72 & 680 & 331 & 2.05 & 1.48 & 44.8 & 93.0 & 90.6 & 79.9 & 13.1 & 65.6 & 359 & 10.9 & 15.9 & 13.5\\

74 & 770 & 390 & 1.97 & 1.48 & 73.2 & 96.2 & 94.0 & 84.7 & 11.5 & 47.9 & 187 & 84.5 &
158 & 122
\end{tabular}
\end{ruledtabular}
\end{small}
\end{table*}

\begin{table*}[!]
\begin{small}
\begin{singlespace}
\caption{Measured growth rates ($10^{-4}/\textrm{s}$) of the ten fastest growing peaks for each experimental run in Table~\ref{table:master} as extracted from the differential colorimetry measurements described in the text. Top row designates the experimental run. Ten entries below each run specify measured growth rates. The minimum ($\sigma^{\textrm{min}}_o$), maximum ($\sigma^{\textrm{max}}_o$) and average ($\sigma^{\textrm{ave}}_o$) values of the growth rate are listed in Table~\ref{table:master}.} \label{table:AllGrowthRates}
\end{singlespace}
\begin{ruledtabular}
\begin{tabular}{c | c | c | c | c | c | c | c | c | c | c | c | c | c | c | c | c | c | c | c}
46 & 47 & 48 & 50 & 51 & 52 & 53 & 54 & 56 & 58 & 60 & 61 & 62 & 63 & 64 & 66 & 69 & 71 & 72 & 74\\ \hline

5.89 & 5.63 & 4.22 & 6.54 & 5.94 & 17.9 & 3.46 & 4.91 & 3.46 & 5.51 & 5.65 & 5.02 & 10.3 & 8.14 & 14.0 & 1.38 & 2.49 & 20.5 & 13.0 & 106\\

5.18 & 2.94 & 5.39 & 6.26 & 2.70 & 13.3 & 3.56 & 7.77 & 4.57 & 3.03 & 2.91 & 2.16 & 9.68 & 9.50 & 10.4 & 0.756 & 2.29 & 17.9 & 15.9 & 114\\

4.43 & 3.33 & 5.04 & 7.85 & 6.21 & 11.4 & 3.76 & 8.24 & 2.39 & 4.36 & 3.00 & 3.51 & 8.49 & 12.4 & 13.5 & 0.888 & 1.87 & 15.3 & 15.0 & 131\\

5.53 & 4.84 & 5.68 & 6.23 & 7.95 & 12.2 & 2.87 & 7.08 & 3.06 & 3.38 & 5.17 & 4.45 & 8.45 & 5.56 & 23.8 & 1.24 & 2.09 & 17.8 & 12.3 & 131\\

5.65 & 2.99 & 4.50 & 8.13 & 8.69 & 15.1 & 3.94 & 10.7 & 3.76 & 3.00 & 3.45 & 4.38 & 12.1 & 6.31 & 17.2 & 1.07 & 1.62 & 13.4 & 13.9 & 84.5\\

5.03 & 3.69 & 4.53 & 4.99 & 7.84 & 13.2 & 3.75 & 5.86 & 3.39 & 4.05 & 3.03 & 4.74 & 10.5 & 7.59 & 17.9 & 0.962 & 1.63 & 14.1 & 15.5 & 151\\

4.48 & 7.79 & 4.79 & 6.29 & 7.56 & 12.2 & 4.73 & 9.75 & 4.68 & 3.82 & 3.88 & 5.28 & 10.8 & 10.6 & 16.1 & 0.848 & 2.03 & 21.0 & 11.8 & 112\\

4.11 & 2.99 & 5.46 & 6.89 & 6.52 & 14.4 & 3.45 & 9.96 & 2.96 & 1.83 & 3.42 & 3.63 & 10.2 & 12.8 & 11.9 & 1.25 & 1.85 & 21.9 & 13.7 & 136\\

5.03 & 4.43 & 4.39 & 5.45 & 8.74 & 12.3 & 5.23 & 4.73 & 3.73 & 2.96 & 3.14 & 4.86 & 7.03 & 5.09 & 13.6 & 0.345 & 2.03 & 12.1 & 10.9 & 158\\

4.97 & 3.19 & 3.84 & 7.17 & 7.94 & 9.45 & 3.15 & 6.35 & 2.95 & 4.07 & 4.23 & 6.10 & 10.8 & 8.55 & 12.2 & 1.00 & 2.03 & 22.5 & 12.9 & 99.0
\end{tabular}
\end{ruledtabular}
\end{small}
\end{table*}

\section{Conclusion}
\label{sec:discussion}
In this work, we implemented a differential colorimetry technique to quantify the growth rate of fluctuations in liquid nanofilms and compared these measurements to model predictions \cite{Dietzel:prl2009,Dietzel:jap2010} which attribute these formations to a  long wavelength thermocapillary instability. At early times, the spacing and growth rate of fluctuations in film thickness is controlled by the competition between large destabilizing thermocapillary stresses and large stabilizing capillary stresses. Linear stability analysis offers an explicit dispersion relation for the rate of growth of the fastest growing unstable mode as a function of geometric parameters and material constants. The measurements reported here were somewhat challenging to carry out since the fluctuation amplitudes within the linear regime spanned only about 20 - 200 nm. The peak reconstruction process could therefore not be obtained by white light or laser  interferometry. Instead, extraction of the growth rate of liquid peak heights relied on reflectance measurements from the air/liquid interface as it deformed in time. The peak reconstruction process incorporated numerous factors affecting the overall reflectance including the spectral response of the camera and illumination source, variations between images due to camera brightness and white balance settings, and small lateral drifts in peak positions caused by slightly non-parallel substrates or other artifacts.

The measured growth rates show very good quantitative agreement with the predictions of the thermocapillary model. This agreement lends further support to the hypothesis that fluctuation growth in such systems is triggered by a long wavelength instability in which thermocapillary stresses are sufficiently strong to pull fluid elongations out-of plane toward the colder proximate substrate. While this study helps further establish the  correlation between the instability growth rate and the normalized wavelength of the fastest growing mode $\Lambda_o = \lambda_o/h_o$ and normalized gap ratio $D = d_o/h_o$, we were unable to test the correlation with $\Delta T$ since the experimental setup could only access the limited range $10 \leq \Delta T \leq 26 \degree \textrm{C}$. Efforts are underway to redesign the system in the hopes of achieving at least one decade variation in $\Delta T$.

Clearly, studies of this type stand to benefit from more data, better statistics and less data scatter but these improvements require ultimately require fabrication of better quality nanofilms free of defects which afford larger measurement areas for analysis. Despite various cleaning and film preparation protocols used over the years, we continue to observe that even uniform and homogeneous polymer nanofilms, once heated, are prone to nucleation and growth of pinholes which then trigger multiple dewetting fronts that interfere with the development and measurement of the intrinsic instability. So far, all experiments have been conducted under ambient conditions but could likely benefit from operation in a clean room. Some studies \cite{Ashley:lang2005} have indicated, however,  that pinholes in heated nanofilms are unavoidable since they originate from a wetting transition that occurs upon heating. While at cooler temperatures the van der Waals potential between the silicon and polystyrene film is attractive in nature (wetting), warmer temperatures can induce repulsion (dewetting). Other studies \cite{Lee:jcp2004} have also demonstrated that residual solvent (toluene) in spun coat nanofilms can lead to local pockets of the film undergoing evaporation which ultimately causes pinhole formation during heating at those locations. Further work is currently underway to try and overcome these difficulties.

\begin{acknowledgments}
We wish to acknowledge financial support from a 2013 NASA Space Technology Research Fellowship (KRF) and the National Science Foundation (EML and SMT). We are also grateful for instrumentation support received from the Molecular Materials Research Center of the Beckman Institute of the California Institute of Technology.
\end{acknowledgments}


%

\end{document}